\documentclass[aps,pre,onecolumn,groupedaddress,times]{revtex4-2}
\usepackage[utf8]{inputenc}
\usepackage{graphics,graphicx,amssymb,amsmath,dcolumn,bm,url,amsbsy,pgf}
\usepackage{booktabs,multirow}
\usepackage{geometry}
\usepackage{physics}
\usepackage{booktabs}
\usepackage[utf8]{inputenc}
\usepackage[T1]{fontenc}
\usepackage{indentfirst}
\usepackage{CJK}
\usepackage{url}
\usepackage{appendix}

\date{}

\begin{document}
\title{Isochronous bifurcations dependence on the driving mode phase shift in two-harmonic standard maps}

\author{Michele Mugnaine}
\email[]{mmugnaine@gmail.com}
\affiliation{Institute of Physics, University of São Paulo, 05508-900 São Paulo, SP, Brazil}
\author{Ricardo L. Viana}
\affiliation{Department of Physics, Federal University of Paran\'a, Curitiba, PR, 81531-980, Brazil}
\affiliation{Interdisciplinary Center for Science, {Technology, and Innovation}, Curitiba, PR, 81530-000, Brazil\relax}
\author{A. M. Ozorio de Almeida}
\affiliation{Brazilian Center for Research in Physics, Rio de Janeiro, RJ, 22290-180, Brazil}
\author{Yves Elskens}
\affiliation{Aix-Marseille Université, CNRS, UMR 7345 PIIM, F-13397, Marseille cedex 13, France}
\author{Iber\^e L. Caldas}%
\affiliation{Institute of Physics, University of São Paulo, 05508-900 São Paulo, SP, Brazil \relax}
\date{\today}

 \begin{abstract}
Some dynamical properties of nonlinear coupled systems can be described by the two-harmonic standard map, a two-dimensional area-preserving system with two parameters, where two distinct arbitrary resonant modes compete. Usually, the initial phase of the resonant modes is considered to be null. In this paper, we consider a non-null phase shift between the two competing isochronous modes that form the system. We observe that a non-zero phase shift alters the phase space, changing the stability and positions of the fixed points. Furthermore, the phase shift can change the dominant mode and create intermediate modes between the main ones. Lastly, we analyze the effect of the phase shift on the onset of secondary shearless curves in the phase space. Thus, different phase shifts result in various scenarios in which secondary shearless curves emerge in the phase space.
 \end{abstract}
\maketitle

\section{Introduction}

Nonlinear coupling has been investigated in several physical systems \cite{sousa2018}, including wave coupling in plasma physics \cite{sagdeev1969,ritz1986,horton2012}, coupled lasers \cite{wiesenfeld1990,kozyreff2000}, and others. The properties of coupled systems depend on the coupling and the energy exchanges among the subsystems \cite{ritz1986,horton2012,ford1961,jackson1963,gendelman2001}. 

A suitable description of coupling properties of non-integrable Hamiltonian systems can be made by the Chirikov-Taylor map \cite{chirikov1971,chirikov1979}, also called standard map, which is an area-preserving map that can be obtained, for example, from studies of kicked oscillators and particles in a magnetic trap \cite{chirikov1987}. In fact, the standard map describes {a typical} oscillating system near a nonlinear resonance  \cite{izraelev1980}.


A generalized version of the standard map can be derived from a Hamiltonian perturbed by a sequence of kicks applied to a {superposition} of waves \cite{baesens1994,cetin2022}. From this Hamiltonian, we can obtain a generalized standard map with a {sum} of resonant modes.  The collection of all modes forms a potential that can be interpreted as a multiple-well potential \cite{baesens1994,cetin2022}.

A Hamiltonian with perturbing kicks can be obtained if we consider a charged particle in a broadband spectrum, \textit{i.e}, a particle of unit mass in an infinite set of electrostatic waves having the same amplitudes, same wave numbers, integer frequencies and zero initial phases \cite{benisti1998,elskens2010}. In Ref. \cite{benisti1998}, Bénisti and Escande studied nonstandard diffusion properties of the system obtained by such Hamiltonian. Furthermore, they consider a finite number of waves and random distinct initial phases for each wave. Considering only one wave, Carlo and coauthors proposed, in Ref \cite{carlo2006}, a Hamiltonian for cold atoms, or a Bose condensate, subjected to a far-detuned standing wave, named atom-optical kicked rotor problem. {Also for} only one wave, it is possible to apply the Hamiltonian with kicks to study the dynamics transverse to the magnetic field of a  relativistic particle moving in a uniform magnetic field and perturbed by a standing electrostatic wave, as shown in Ref. \cite{sousa2013}.

Recently, we analyzed a map, that can be obtained for only two terms of the summation and different amplitudes for each term, named two-harmonic standard map \cite{mugnaine2024}. Such {a} map was proposed with the purpose of presenting a simple system which simulates the competition between two isochronous resonant modes and exhibits isochronous bifurcations. An isochronous bifurcation is defined as the emergence of  distinct chains of periodic islands with the same frequency in the same region of phase space. These islands are also named isochronous and they emerge in twist systems as a response to the competition between two arbitrary resonant modes.

{Due to violating the twist condition (named a non-degeneracy condition for Hamiltonian continuum systems), different dynamical phenomena emerge in the phase space. {The shearless curve,  for which the derivative of the rotation number {with respect to} the action vanishes, is an example of nontwist phenomena that occur because of the non-monotonicity of the rotation number \cite{morrison2000,del2000}. Furthermore, shearless curves are robust in the sense that they can survive the destruction of neighbor invariant curves on both sides with respect to the action variable.} {Moreover}, Dullin et al. \cite{dullin2000} demonstrated the existence of shearless (twistless) curves in the twist conservative Hénon map. In this case, they related the shearless curve to an extreme value for the internal rotation number, and since the shearless curve is inside an island, it is called a secondary shearless curve. Abud and Caldas also identified these secondary shearless curves in the standard twist map \cite{abud2012}. {Most recently}, other examples of secondary shearless curves have been found \cite{leal2025}}

In our studies developed in Refs. \cite{mugnaine2024} and \cite{leal2025} we considered that, as for the standard map, the phase is null for all waves/resonant terms. The same {assumption} was made for the analysis of the extended standard map, analyzed in Refs. \cite{ketoja1989,greene1990,baesens1994}. However, it is possible to consider different phases for the system. Frahm and Shepelyansky considered random phase shift angles in the standard map and they observed a random behavior for small time scales \cite{frahm2009}. Phase shifted kicks were also considered in the standard map by Cavallasca, Artuso and Casati and the consequence is symmetry breaking and the emergence of ratchet current in the transport of chaotic orbits \cite{cavallasca2007}.  Distinct phases can also be considered in the wave-particle problem {, resulting in various types of transport similar to the effect of a noise} \cite{benisti1998,elskens2010,escande2002, elskens2007,elskens2012}.

In {this} work, we consider distinct phases for the two-harmonic standard map and analyze the effect of a phase shift in the system. We investigate the impact of phase shifts equal to $\varphi=\pi$ and $\varphi=\pi/2$ in the phase space, in isochronous bifurcations and in the emergence of secondary shearless curves. From our results, we show that the phase shift has an important role in the positions of elliptic points, which changes the position of
periodic islands in phase space. We also verify that the phase shift modifies the isochronous bifurcations, altering the type of bifurcation and the intermediate modes between the dominant ones. As a last consequence, we identify the onset of internal shearless curves due to different phase shifts.
 
This paper is organized as follows: the studied model is presented in Sec. \ref{secII} and the impact of the phase shift in the modes of the system {is} discussed in Sec. \ref{secIII}. The transition by isochronous bifurcations is studied in Sec. \ref{secIV}. Our analysis about secondary shearless curve is presented in Sec. \ref{secV}. Our conclusions
are provided in the last section.

\section{The model}
\label{secII}
A generalized version of the standard map {can be} described by the {time-dependent Hamiltonian}, {based on Ref.} \cite{cetin2022},
{
\begin{eqnarray}
    H=\dfrac{y^2}{2}-K\left[\sum_{j=1}^W \dfrac{1}{4\pi^2 j²} \cos(2\pi j x)\right]\sum_{n=-\infty}^{\infty} \delta(t-n\tau)
    \label{hamiltonian}
\end{eqnarray}}
where {parameter} $K$ controls the integrability of the system{; $K\ne0$ indicates a non-integrable system}. The parameter $W$ is an integer, and the periodic kicks are modeled by the periodic Dirac $\delta$ {distribution}. {We set $\tau = 1$ with no loss of generality.} The Hamiltonian associated with the standard map is recovered when $W=1$. From the Hamiltonian (\ref{hamiltonian}), we obtain the following generalized standard map
\begin{eqnarray}
    \begin{aligned}
        x_{n+1}&=x_n+y_{n+1}\\
        y_{n+1}&=y_n-\sum_j\dfrac{K}{2\pi j}  \sin(2\pi j x_n)    
    \end{aligned}
    \label{gsm}
\end{eqnarray}
where $x$ and $y$ {can be} taken modulo 1. Each term in the summation can be taken as a resonant perturbation mode, and the collection of all modes forms a potential that can be interpreted as a multiple-well potential \cite{cetin2022}.

{The two-harmonic standard map \cite{mugnaine2024}, considered in this work, is an extension of (\ref{gsm}) where one considers two frequencies $m_1$ and $m_2$ and independent amplitudes $K_1$ and $K_2$}. Furthermore, adding a phase shift $\varphi$ on the argument of the second harmonic of the two-harmonic standard map \cite{mugnaine2024}, we obtain the equations
\begin{eqnarray}
    \begin{aligned}
        x_{n+1}&=x_n+y_{n+1},\\
        y_{n+1}&=y_n-\dfrac{K_1}{2\pi m_1}\sin(2 \pi m_1 x_n) - \dfrac{K_2}{2 \pi m_2} \sin(2 \pi m_2 x_n + \varphi),
    \end{aligned}
    \label{eq:ESNM}
\end{eqnarray}
where $K_1, K_2 \in \mathbb{R}$ and $m_1, m_2 \in \mathbb{N}$. The numbers $m_1$ and $m_2$ identify the modes of the system, {\textit{i.e.} the number of elliptic points at $y=0$,} and, depending on the amplitudes $K_1$ and $K_2$, the system can exhibit $m_1$ to $m_2$ islands. {In this paper, we always consider $m_2 > m_1$ and values of $K_1$ and $K_2$ in the range [0,4]}. { We consider (\ref{eq:ESNM}) on the unit torus, \textit{i.e.},} we take $\mod$ 1 for both variables in (\ref{eq:ESNM}).

\begin{figure}[!h]
	\begin{center}
		\includegraphics[width=1.0\textwidth]{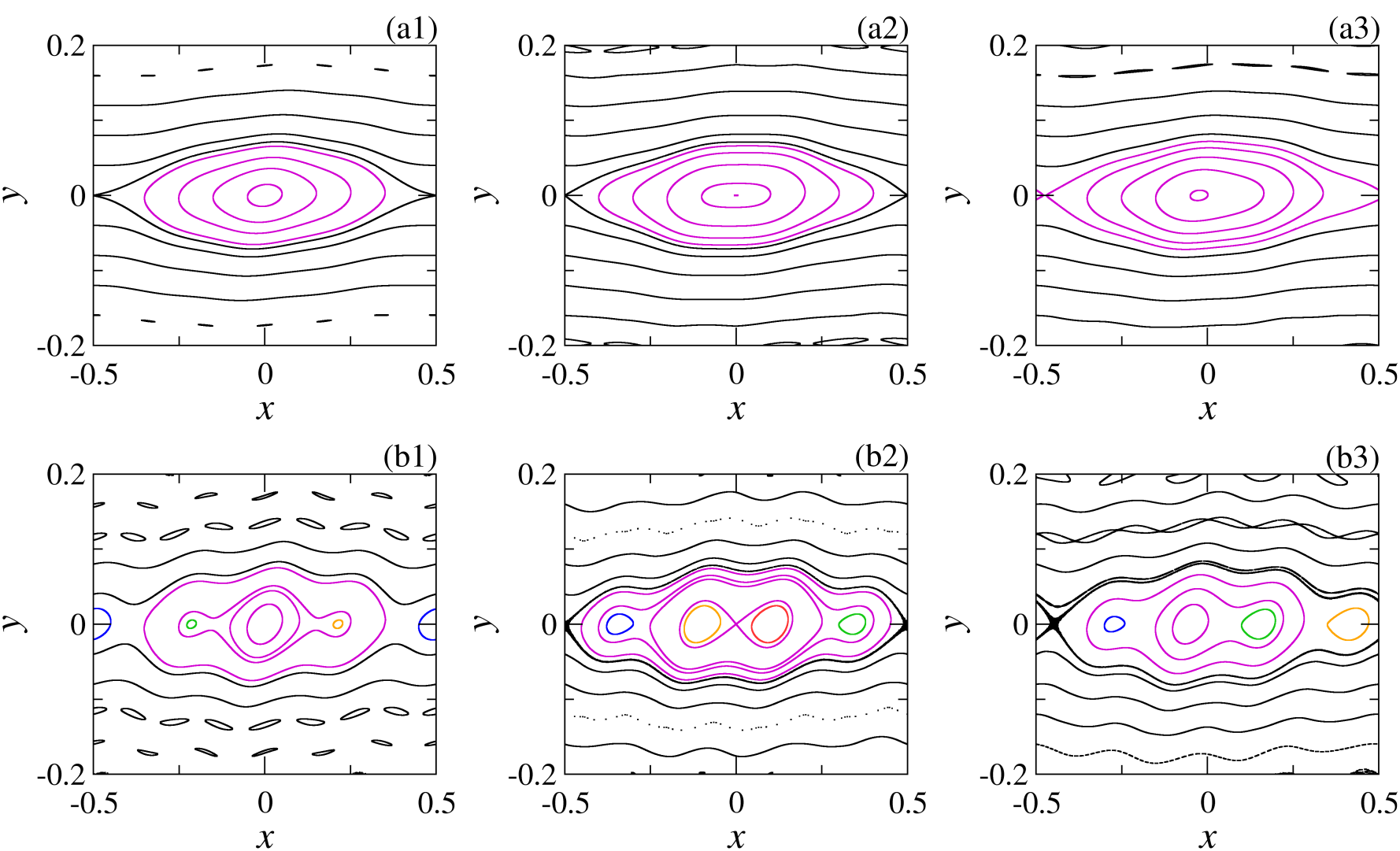}		
		\caption{The impact of the phase $\varphi$ for the two-harmonic standard map for $m_1=1$ and $m_2=4$, with $K_2=0.04$ (first row) and $K_2=0.25$ (second row). The phase for each column is different: $\varphi=0.0$ for the first column, $\varphi=\pi$ for the second column and $\varphi=\pi/2$ for the third one. For all phase spaces, $K_1=0.05$.}
		\label{fig1}
	\end{center}
\end{figure}

With $\varphi=0$, we recover the two-harmonic standard map, analyzed in Ref. \cite{mugnaine2024}. In this work, {our objective} is to analyze the role of a nonzero phase {shift} $\varphi$ in the system and its impact on the number of islands and on isochronous bifurcations, \textit{i.e.}, routes from mode $m_1$ to mode $m_2$. For this analysis, we compute the phase portraits for different values of $\varphi$. Specifically, we chose $\varphi=0$ (the {original} two-harmonic standard map), $\varphi=\pi$, and $\varphi=\pi/2$ (representing maps with a phase shift). For the modes, we select $m_1=1$ and $m_2=4$. The phase portraits for the three values of $\varphi$ and the two modes are shown in Figure \ref{fig1}.

In Figures \ref{fig1} (a1) and (b1), we present the phase spaces for $\varphi=0$ with different amplitudes of $K_2$. When $K_1=K_2=0.05$, \textit{i.e.}, panel (a1), there is only one island around the elliptic point at $(0,0)$, indicating the predominance of the mode $m_1=1$. When $K_2$ increases to $K_2=0.25$, the configuration shown in Figure \ref{fig1} (b1) emerges, where four islands of period 1 are observed around four distinct elliptic points, representing the mode $m_2=4$.

Similar scenarios occur for the phases $\varphi=\pi$ (second column) and $\varphi=\pi/2$ (third column): the mode $m_1=1$ is predominant in the panels labeled (a), while the mode $m_2=4$ is predominant in the panels labeled (b). However, examining the phase spaces reveals that the phase $\varphi$ affects the elliptic points by altering their positions and/or stability. For example, at $\varphi=0$, there are two elliptic points at positions $x=0$ and $x=0.5$. At $\varphi=\pi$, these fixed points are hyperbolic, while at $\varphi=\pi/2$, there are no elliptic points at these positions. For $\varphi=\pi$, the second harmonic is still a sine function,  but negative; thus, $x=0$ and $x=0.5$ are still fixed points for any value of $K_1$, $K_2$, $m_1$ and $m_2$. For $\varphi=\pi/2$, the second harmonic is a cosine, so the fixed point position depends on all parameters of the system. These analytical results can be found in the appendix.

\section{Phase shifting and the modes of the system}
\label{secIII}
As observed in Ref. \cite{mugnaine2024}, intermediate modes may appear on the route from mode $m_1$ to mode $m_2$. Thus, the number of elliptic points on the line $y=0$ depends on all parameters of the system. We compute the number of elliptic points for different values of $K_1$ and $K_2$ and for all combinations of $m_1$ and $m_2$ with $m_1=1,2,...,5$ and $m_2\in[m_1+1,6]$. Here, we present four different combinations of $(m_1,m_2)$ that represent the general results:  (1,4), (1,5), (2,6) and (4,5). All parameter spaces can be found in the Supplementary Material \cite{sm}.

By searching for fixed points of period 1 on the line $y=0$, we compute the number of distinct elliptic points. We present the parameter spaces $K_1 \times K_2$, where the color indicates the number of distinct elliptic points.  Additionally, the number of elliptic points is labeled by numbers in the corresponding colored regions.

\begin{figure}[!h]
	\begin{center}
		\includegraphics[width=1.0\textwidth]{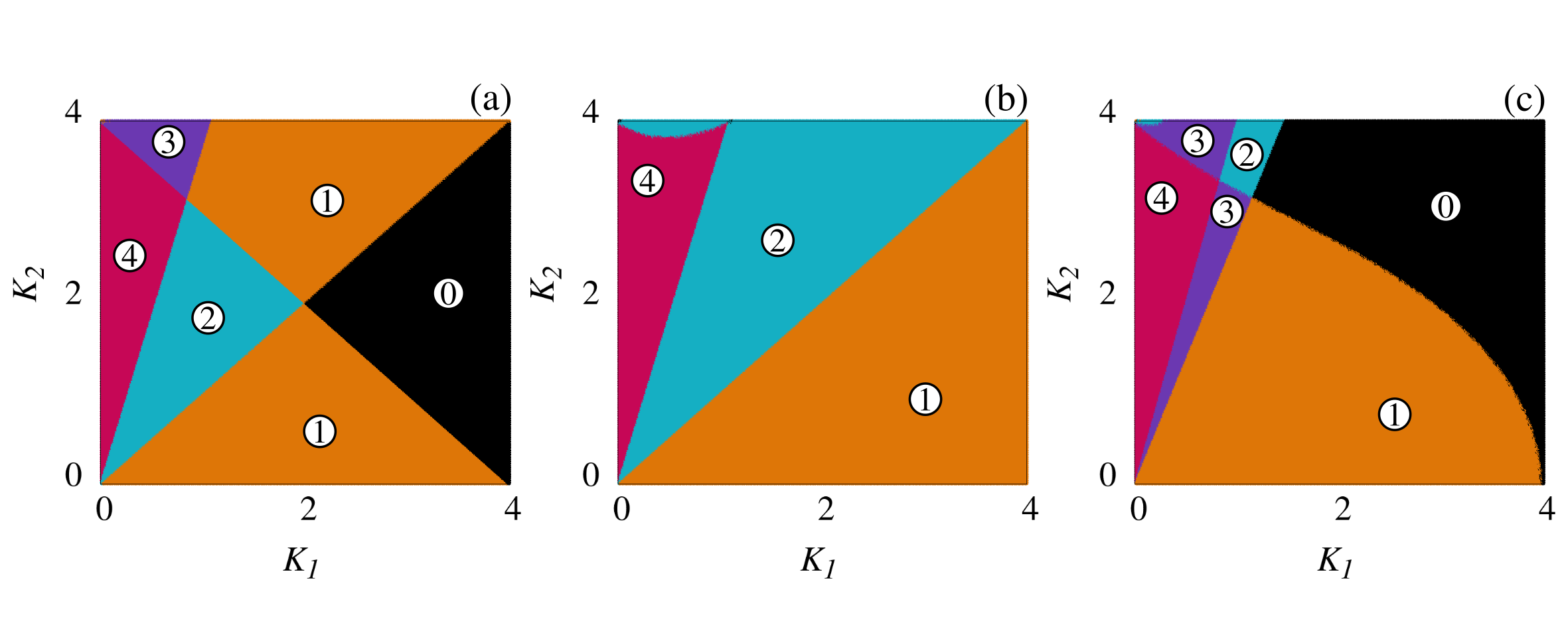}		
		\caption{Parameter spaces for the number of elliptic points for $m_1=1$, $m_2=4$. The numbers indicate the number of elliptic points for the corresponding colored region. The value of $\varphi$ for each case is (a) $\varphi=0$, (b) $\varphi=\pi$ and (c) $\varphi=\pi/2$. }
		\label{fig2}
	\end{center}
\end{figure}

Firstly, we present the parameter spaces for $m_1=1$ and $m_2=4$ and the three chosen values of $\varphi$. As shown in a previous work \cite{mugnaine2024}, there is an intermediate mode $m=2$ {on the way} from one to four elliptic points. This result can be checked in Figure \ref{fig2} (a), {where} $\varphi=0$. {In this configuration,}  the {bifurcation} from mode 1 to mode 2 occurs when $K_2=K_1$, with $K_1\in[0,2]$. When $K_2=3.65 K_1$, the mode 4 emerges in the system {for $0<K_1\le 0.85$}.

For $\varphi=\pi$ and $\varphi=\pi/2$, the parameter spaces are shown in Figure \ref{fig2}(b) and \ref{fig2}(c), respectively. We observe that, for both cases, there is also an intermediate mode between mode 1 and mode 4. However, while the intermediate mode is 2 for $\varphi=\pi$, it is $m=3$ for $\varphi=\pi/2$. The bifurcation lines $K_2=K_1$ and $K_2=3.65K_1$ are the same for $\varphi=0$ and $\varphi=\pi$, but the interval of $K_1$ where the bifurcation occurs differs for each $\varphi$. For $\varphi=\pi/2$, we find $K_2=2.7K_1$, between the regions of modes 1 and 3, and $K_2=4 K_1$ for the $3\to 4$ transition. The first general result is that the different phases $\varphi$ influence the intermediate modes, either by changing the mode itself or altering the interval in which the transition occurs. This result is also found {with} the pair $(m_1=1,m_2=6)$.

\begin{figure}[!h]
	\begin{center}
		\includegraphics[width=1.0\textwidth]{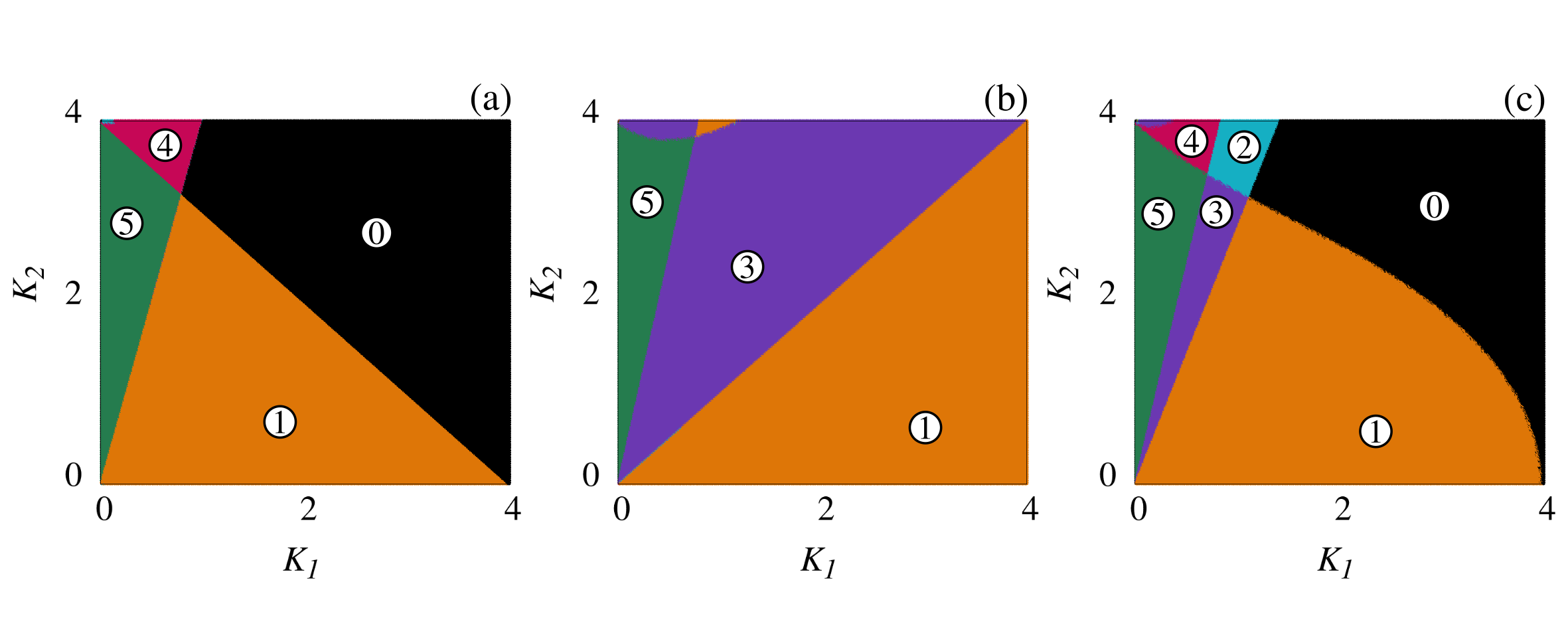}		
		\caption{Emergence of intermediate modes for $\varphi\ne 0$. The parameter spaces indicate the  number of elliptic points for $m_1=1$, $m_2=5$. The phases are (a) $\varphi=0$, (b) $\varphi=\pi$ and (c) $\varphi=\pi/2$. }
		\label{fig3}
	\end{center}
\end{figure}

Next, we choose the modes $m_1=1$ and $m_2=5$ to represent the second general result: the emergence of intermediate modes. In Figure \ref{fig3} we have the respective parameter spaces for (a) $\varphi=0$, (b) $\varphi=\pi$ and (c) $\varphi=\pi/2$.

In Figure \ref{fig3} (a), we observe the parameter space for $\varphi=0$ with no intermediate mode for $1\to5$ transition. The transition occurs when $K_2=4K_1$ for $K_1 \in (0,0.8]$. When $\varphi=\pi$, we have the parameter space shown in Figure \ref{fig3} (b). In this space, we observe an intermediate mode, the mode 3. The same intermediate mode is observed in Figure \ref{fig3} (c), where $\varphi=\pi/2$. The $1\to3$ transition occurs when $K_2=K_1$, for $\varphi=\pi$ and {any} value of $K_1\in(0,4]$. {For $\varphi=\pi/2$, the transition occurs} when $K_2=2.8K_1$ and $K_1\in(0,1.13]$. The second transition, $3 \to 5$, occurs on the line $K_2=5K_1$ when $K_1\in (0,0.76]$ for $\varphi=\pi$, and on the line $K_2=4.8K_1$ when $K_1\in (0,0.72]$ and $\varphi=\pi/2$. Our results suggest that adding a phase shift between the two {harmonics} leads to the emergence of intermediate modes in the system

In contrast to the previous result, there are scenarios where the addition of a non-zero phase does not result in new intermediate modes. However, the phase does influence when the transition occurs, \textit{i.e.}, the bifurcation lines. This occurs for the pairs $(m_1,m_2)$= (1,2), (1,3), (2,4), (2,6), (3,4), (3,5), (3,6), and (5,6). The respective parameter spaces and bifurcation lines can be checked in \cite{sm}.



\begin{figure}[!h]
	\begin{center}
		\includegraphics[width=0.35\textwidth]{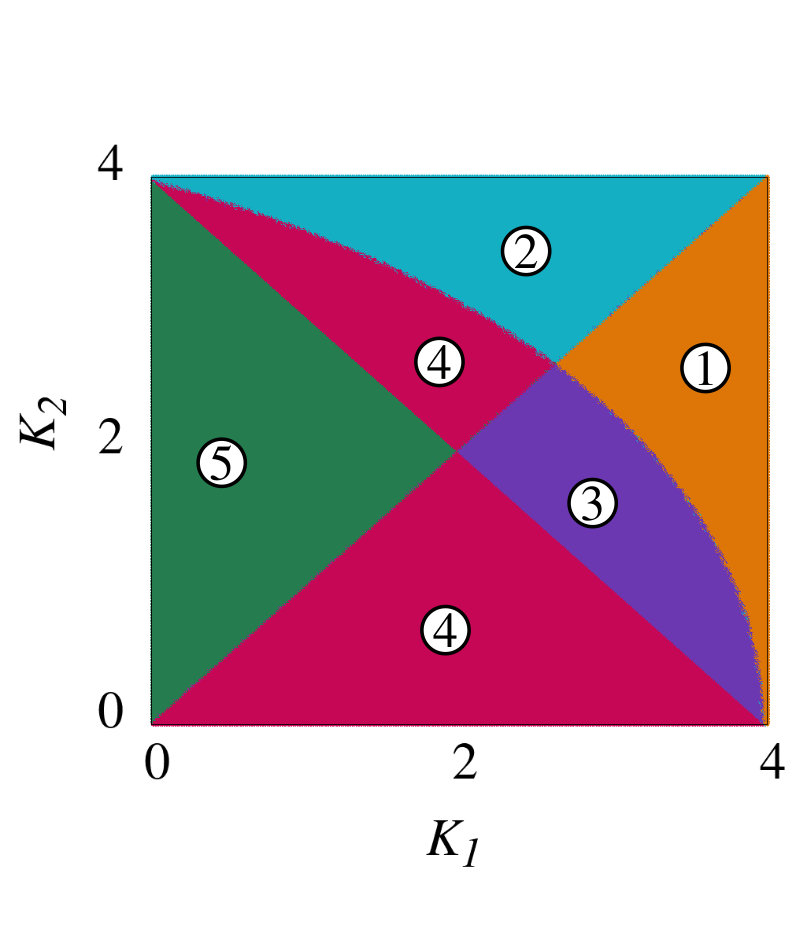}		
		\caption{Invariance of the parameter space for $m_1=4$ and $m_2=5$. The bifurcation curves and the colored region remain identical for $\varphi=0, \pi$ and $\pi/2$.}
		\label{fig5}
	\end{center}
\end{figure}

The final general result concerns the invariance of the bifurcation curves. For certain combinations of $m_1$ and $m_2$, the bifurcation curves remain unchanged {for $\varphi=0, \pi$ and $\pi/2$}, but the colored regions differ. This occurs for $(m_1, m_2) = (2,3), (2,5),$ and $(4,6)$. The corresponding parameter spaces can be found in the Supplementary Material \cite{sm}. Here, we highlight a particular case where both the bifurcation curves and the colored regions remain identical for all three values of $\varphi$. In summary, the parameter space is invariant for the three values of $\varphi$ and it is presented in Figure \ref{fig5}.

For all analyzed values of $\varphi$, the parameter space is the one presented in Figure \ref{fig5} for $m_1=4$ and $m_2=5$. The $4\to5$ transition occurs at the bifurcation curve $K_2=K_1$. This invariance is due to the bifurcation of different fixed points in the system. For $\varphi=0$, the fixed point $(x^*,y^*)=(0.5,0)$ goes through a bifurcation when $K_2=K_1$; for $\varphi=\pi$, the fixed point $(0,0)$ changes its stability for the same bifurcation line; and for $\varphi=\pi/2$, the point $(0.25,0)$ bifurcates when $K_2=K_1$. The mathematical computations are based on the analysis in Appendix.

\section{Transitions by Isochronous Bifurcations}
\label{secIV}
Fixed points of the system can undergo two types of codimension-one bifurcations: pitchfork or saddle-node \cite{mugnaine2024}. The $m_1\to m_2$ transitions can be formed by one or a combination of these bifurcations. In this section, we study the isochronous bifurcations themselves and how they are affected by the addition of non-zero phase $\varphi$. For this, we compute the bifurcation diagrams of the fixed points in relation to the parameter $K_2$. For all the bifurcation diagrams presented, we chose $K_1=0.05$ and the black (gray) lines indicate the elliptic (hyperbolic) points in panels (a) for the next figures. 

\begin{figure}[!h]
	\begin{center}
		\includegraphics[width=1.0\textwidth]{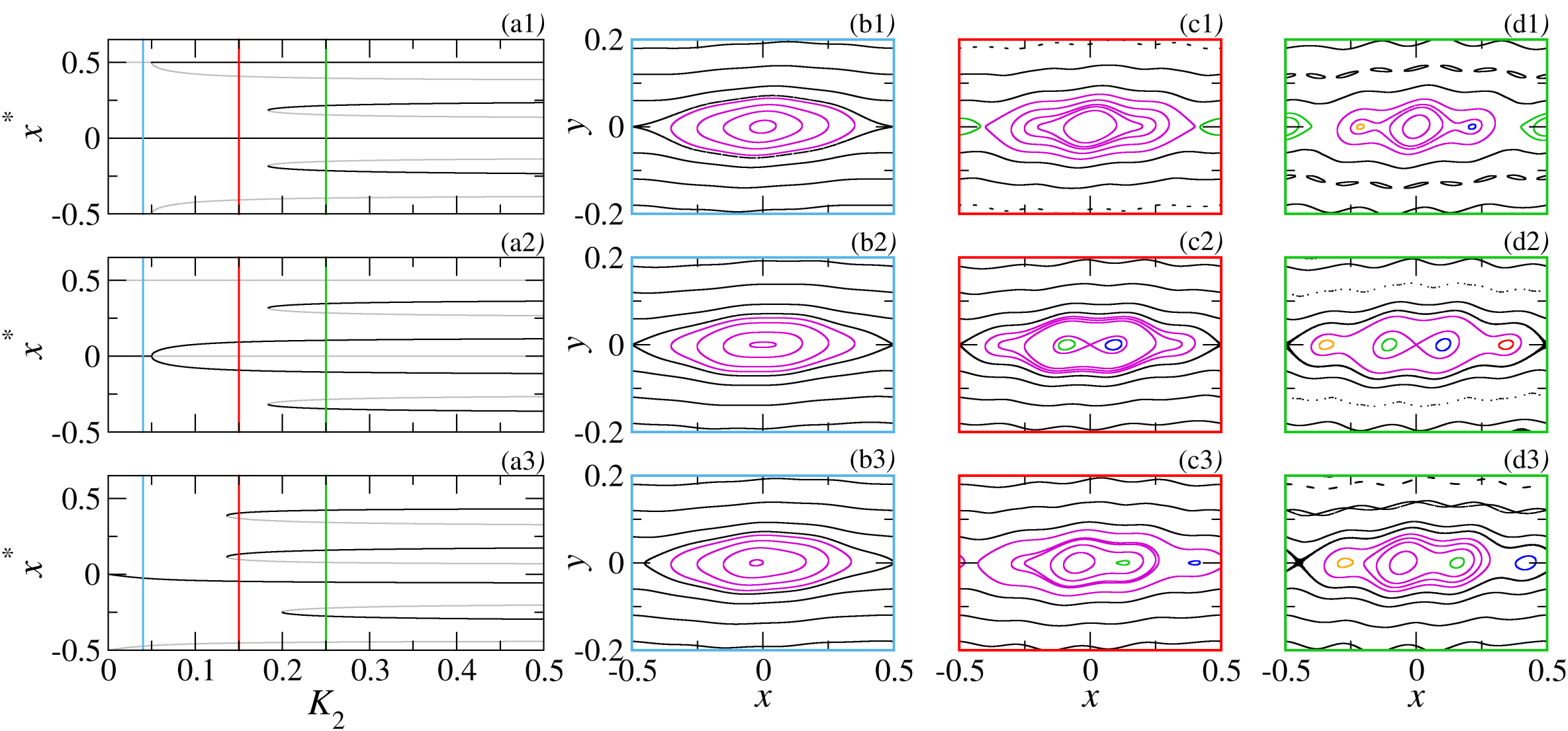}		
		\caption{Isochronous bifurcations for $m_1=1$ and $m_2=4$ {with $K_1=0.05$}. We present the bifurcation diagrams (a) for the fixed points and the phase spaces (b-d) for the parameters $K_2$ indicated by the color lines for $\varphi=0.0$ (first row), $\varphi=\pi$ (second row) and $\varphi=\pi/2$ (third row). }
		\label{fig6}
	\end{center}
\end{figure}

In Figure \ref{fig6} we observe the {route} from $m_1=1$ to $m_2=4$ islands for the three chosen values of $\varphi$. For the two first rows, the {route involves} first a pitchfork bifurcation from 1 to 2 islands and then two saddle-node bifurcations occur simultaneously. The third row, for $\varphi=\pi/2$, {displays} a different scenario: as seen in the previous section, the intermediate mode is 3 and both transitions $1 \to 3$ and $3 \to 4$ occur by saddle-node bifurcations. {In Figure 5, the final modes are the same, but the final four islands sequence in phase space depends on the phase shift.}

\begin{figure}[!h]
	\begin{center}
		\includegraphics[width=1.0\textwidth]{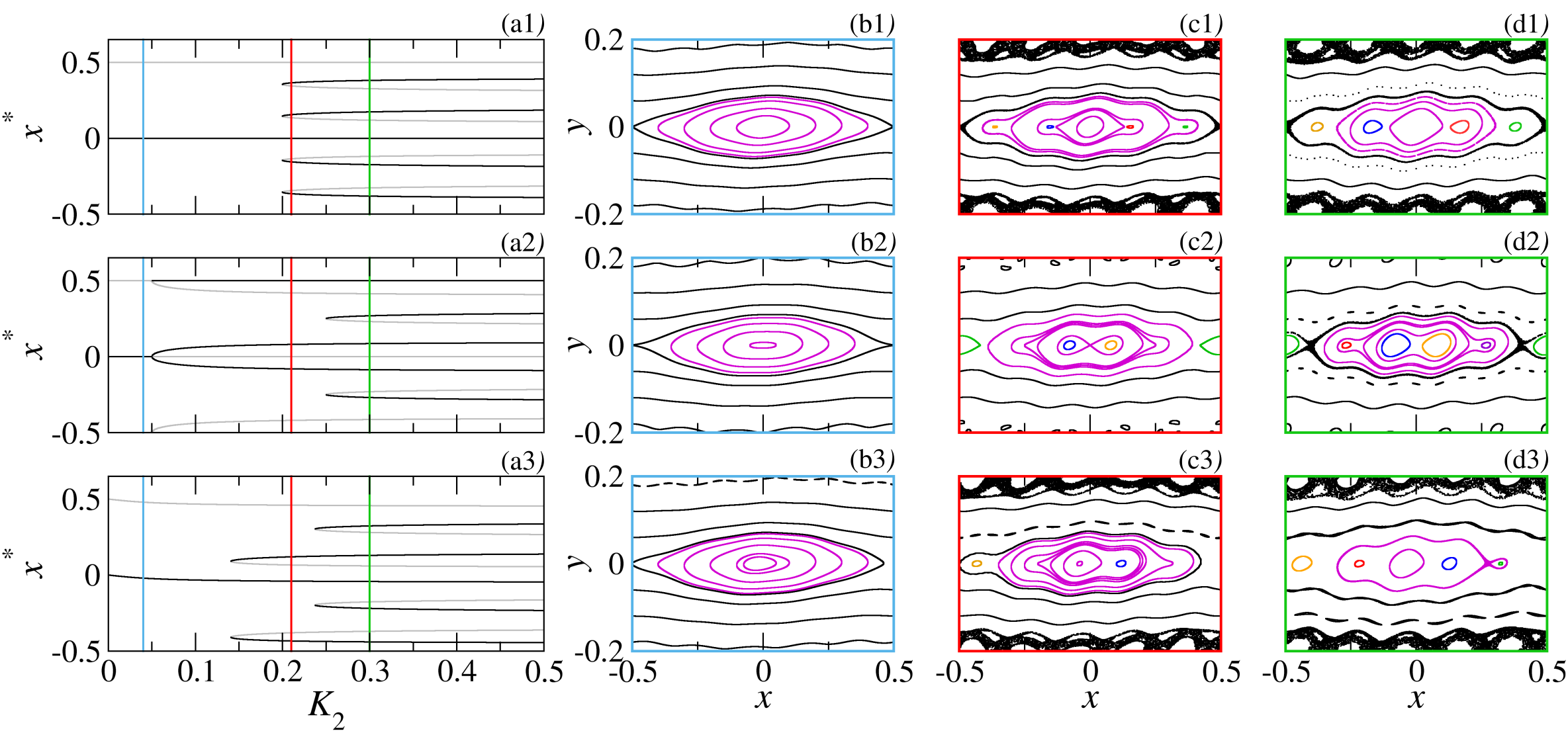}		
		\caption{Route from $m_1=1$ to $m_2=5$ islands {with $K_1=0.05$}. Bifurcation diagrams (a) for the fixed points and phase spaces (b-d) for the parameter $K_2$ indicated by the color lines for $\varphi=0.0$ (first row), $\varphi=\pi$ (second row) and $\varphi=\pi/2$ (third row).}
		\label{fig7}
	\end{center}
\end{figure}

Next, we consider the case where $m_1=1$ and $m_2=5$, where the inclusion of a non-zero phase is responsible for the emergence of an intermediate mode. The results are shown in Figure \ref{fig7}. In the first row, we have $\varphi=0$ and the $1\to5$ transition occurs by four saddle-node bifurcations, which occur at the same value of $K_2$. For $\varphi=\pi$ we have the {sequence} shown in the second row, where two pitchfork bifurcations in $K_2=K_1$ are responsible for the $1\to3$ transition. On increasing the value of $K_2$, two saddle-node bifurcations occur and we observe the mode $m_2=5$. The third row represents the case where $\varphi=\pi/2$: in this case, we only have saddle-node bifurcations for both $1\to 3$ and $3\to 5$ transitions. {In Figure 6, the final configurations are also different for the considered three phase shift values.}

A similar scenario occurs for the $2 \to 6$ transition: only saddle-node bifurcations for $\varphi=0$ and $\pi/2$, and pitchfork bifurcations for $\varphi=\pi$. {The bifurcation diagrams for this route are available in} the Supplementary Material \cite{sm}.

\begin{figure}[!h]
	\begin{center}
		\includegraphics[width=0.9\textwidth]{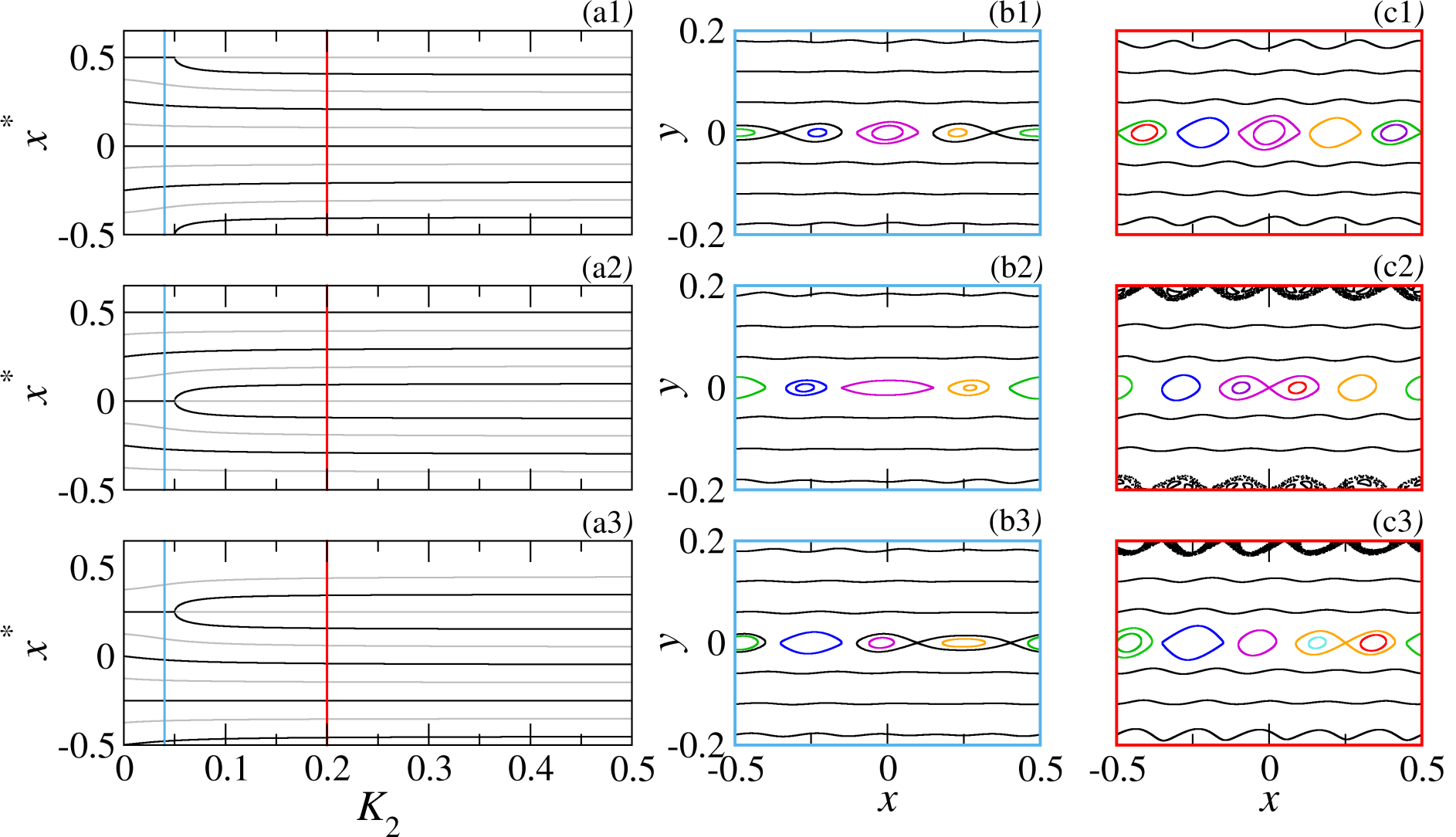}		
		\caption{Isochronous bifurcations for $m_1=4$, $m_2=5$ {and $K_1=0.05$}. Bifurcation diagrams (a) for the fixed points and phase spaces (b-d) for the parameter $K_2$ indicated by the color lines for $\varphi=0.0$ (first row), $\varphi=\pi$ (second row) and $\varphi=\pi/2$ (third row).}
		\label{fig8}
	\end{center}
\end{figure}

Lastly, we analyze the case  $m_1=4$ and $m_2=5$ where the bifurcation lines along with the parameter spaces are equal for all values of $\varphi$. In Figure \ref{fig8}, we observe the bifurcation diagrams in panels (a) and the corresponding phase spaces in panels (b) and (c).

The bifurcation diagrams {show} that bifurcations occur for the same values of $K_2$ but in different fixed points. For $\varphi=0$ (the first row), the $4\to5$ transition occurs by a pitchfork bifurcation in the fixed point $x^*=0.5$. For the second row, we have  $\varphi=\pi$ and also a pitchfork bifurcation, but this time, in the fixed point $x^*=0$. Lastly, a pitchfork bifurcation occurs in the fixed point $x^*=0.25$ for $\varphi=\pi/2$, the sequence shown in the third row. With these results, we have a scenario where the addition of a non-zero phase changes the fixed point which goes through a bifurcation, but the bifurcations are the same (pitchfork bifurcations) and they occur at the same value $K_2$.

{Finally}, we analyze all the routes from mode $m_1$ to mode $m_2$ for the three values chosen from $\varphi$. The routes are shown in Table \ref{tab1}. The letters $P$ and $SN$ indicate the occurrence of pitchfork and saddle-node bifurcations, respectively.
\begin{table}[]
    \centering
\begin{tabular}{ |c|c|c|c|} 
		\hline
		Transition & $\varphi=0$ & $\varphi=\pi$ & $\varphi=\frac{\pi}{2}$  \\ 
		\hline
		$1 \to 2$ & $1 \xrightarrow{P} 2$ &$1 \xrightarrow{P} 2$ & $1 \xrightarrow{SN} 2$\\
		$1 \to 3$ & $1 \xrightarrow{SN} 3$ & $1 \xrightarrow{P} 3$ & $1 \xrightarrow{P} 3$\\
		$1 \to 4$ & $1 \xrightarrow{P} 2 \xrightarrow{SN} 4$ & $1 \xrightarrow{P} 2 \xrightarrow{SN} 4$ & $1 \xrightarrow{SN} 3 \xrightarrow{SN} 4$ \\
		$1 \to 5$ & $1 \xrightarrow{SN} 5$ & $1 \xrightarrow{P} 3 \xrightarrow{SN} 5$ & $1 \xrightarrow{SN} 3 \xrightarrow{SN} 5$\\
		$1 \to 6$ & $1 \xrightarrow{P} 2 \xrightarrow{SN} 4 \xrightarrow{SN} 6$ & $1 \xrightarrow{P} 2 \xrightarrow{SN} 4 \xrightarrow{SN} 6 $& $1 \xrightarrow{SN} 3 \xrightarrow{SN} 5 \xrightarrow{SN} 6 $ \\
		$ 2 \to 3$ & $2 \xrightarrow{P} 3$ & $2 \xrightarrow{P} 3$ & $2 \xrightarrow{P} 3$\\
		$2 \to 4$ & $2 \xrightarrow{P} 4$ & $2 \xrightarrow{P} 4$ & $2 \xrightarrow{SN} 4$\\
		$2 \to 5$ & $2 \xrightarrow{P} 3 \xrightarrow{SN} 5$ & $2 \xrightarrow{P} 3 \xrightarrow{SN} 5$ & $2 \xrightarrow{P} 3 \xrightarrow{SN} 5$\\
		$2 \to 6$ & $2 \xrightarrow{SN} 6$ & $2 \xrightarrow{P} 6$& $2 \xrightarrow{SN} 6$\\
		$3 \to 4$ & $3 \xrightarrow{P} 4$& $3 \xrightarrow{P} 4$ & $3 \xrightarrow{SN} 4$\\
		$3 \to 5$ & $3 \xrightarrow{SN} 5$ & $3 \xrightarrow{P} 5$ & $3 \xrightarrow{SN} 5$\\
		$3 \to 6$ & $3 \xrightarrow{P} 6$ & $3 \xrightarrow{P} 6$ & $3 \xrightarrow{SN} 6$\\
		$4 \to 5$ & $4 \xrightarrow{P} 5$ & $4 \xrightarrow{P} 5$ & $4 \xrightarrow{P} 5$\\
		$4 \to 6$ & $4 \xrightarrow{P} 6$ & $4 \xrightarrow{P} 6$ & $4 \xrightarrow{P} 6$\\
		$5 \to 6$ & $5 \xrightarrow{P} 6$ & $5 \xrightarrow{P} 6$ & $5 \xrightarrow{SN} 6$\\
		\hline
	\end{tabular}
    \caption{Summary of the types of bifurcations {for all pairs of harmonics  $1 \leq m_1 < m_2 \leq 6$ and $0 \leq K_1 \leq 4$, $0 \leq K_2 \leq 4$}. The letters $P$ and $SN$ indicate pitchfork and saddle-node bifurcations, respectively.}
    \label{tab1}
\end{table}

From the results shown in Table \ref{tab1}, we observe that the phase $\varphi$ can alter the bifurcation that occurs in each route. This is observed in transitions $1\to 3$, $1 \to 4$, $1 \to 5$, $1 \to 6$, $2 \to 4$, $2 \to 6$, $3 \to 4$, $3\to 5$, $3 \to 6$ and $5 \to 6$. We can observe a predominance of pitchfork bifurcations for $\varphi=\pi$ and of saddle-node bifurcations for $\varphi=\pi/2$. For $\varphi=\pi$, we always have a pitchfork bifurcation in the fixed point $x=0.0$ when $K_2=K_1$ and, consequently, is impossible to have routes with just saddle-node bifurcations as observed for $\varphi=0$ and $\varphi=\pi/2.$

\section{Secondary shearless curves}
\label{secV}
As shown in \cite{leal2025}, the emergence of secondary shearless curves is commonly observed in twist systems with resonant mode coupling. For the two-harmonic standard map, {it was observed} three  patterns for the emergence of shearless curves. The first pattern {was} formed by the emergence of a single shearless curve, before the occurrence of a pitchfork bifurcation. A second pattern involved the formation of shearless curves in pairs {where} these curves appear as corresponding maximum and minimum pairs within the internal rotation profile. Finally, the third pattern represented the emergence of shearless curves in distinct islands.

The identification of secondary shearless curves is performed by the analysis of the internal rotation {(or winding)} number $\omega_{\mathrm{in}}$, inside an island, defined \cite{abud2012} as
\begin{eqnarray}
    \omega_{\mathrm{in}}=\lim_{n\to \infty} \dfrac{1}{2\pi n} \sum_{n=1}^\infty P_n \hat{\theta} P_{n+1}
    \label{eq:win}
\end{eqnarray}
where $P_n \hat{\theta} P_{n+1}$ is the angle between two consecutive points, $P_n$ and $P_{n+1}$, in the phase space. Similar to its global correspondent, the limit in (\ref{eq:win}) converges for periodic and quasi periodic solutions and does can fail to exist for chaotic solutions. A secondary shearless curve is identified by a local extremum point in the internal winding number profile, as shown in details in Refs. \cite{leal2025,abud2012}.

In this section, we analyze the impact of the phase shift $\varphi$ in the emergence of secondary shearless curves. Considering all the combinations of $m_1$ and $m_2$ for $m_1\in[1,5]$ and $m_2\in[m_1+1,6]$, we found three possible scenarios for the impact of the phase shift on the secondary shearless curves.

\begin{figure}[!h]
	\begin{center}
		\includegraphics[width=0.5\textwidth]{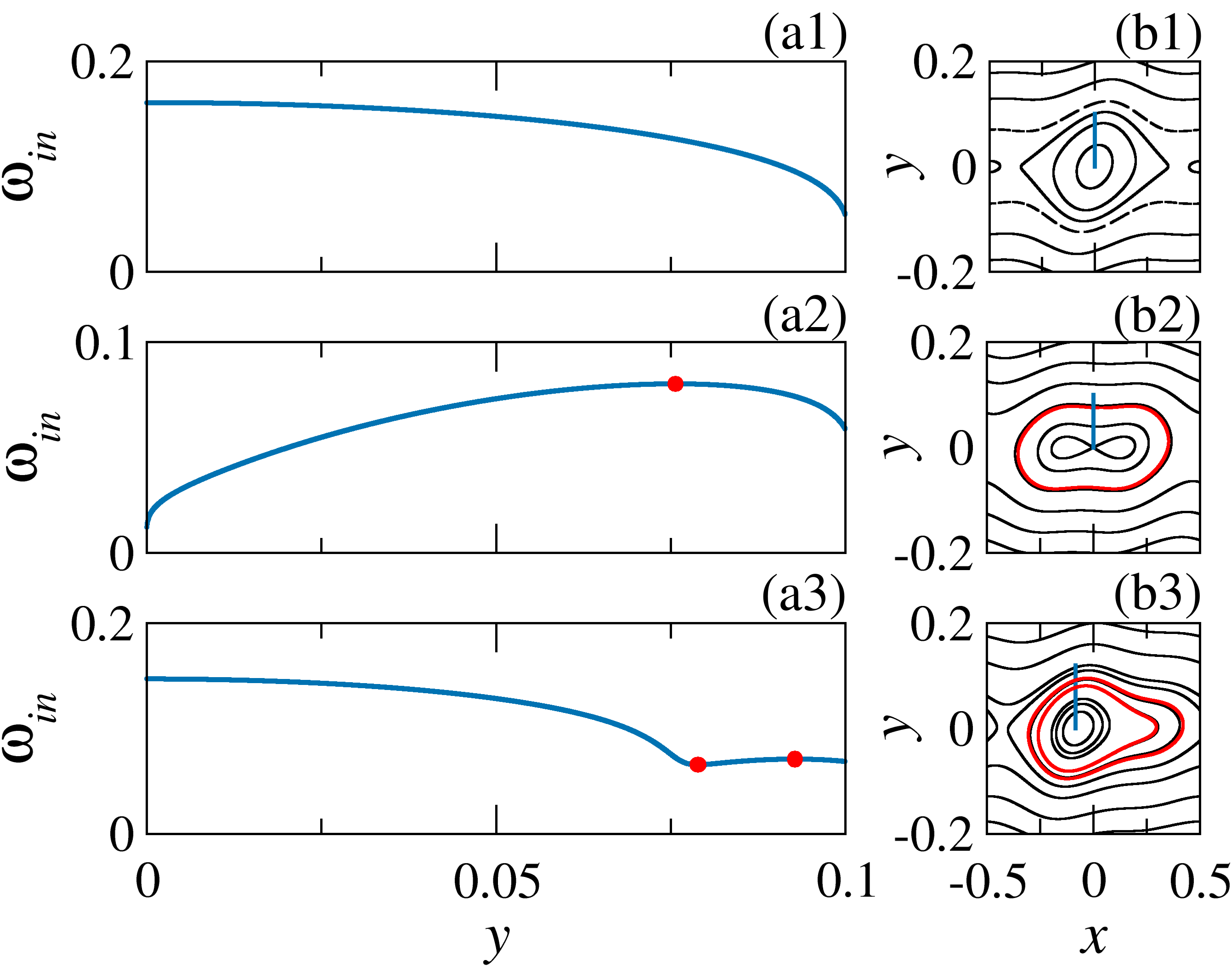}		
		\caption{Emergence of secondary shearless curves when $\varphi\ne 0$, for $m_1=1, m_2=2$, $K_1=0.1$ and $K_2=0.15$. In panels (a) we show the internal winding number profile with the extremum points highlighted by the red dots. The respective phase portraits are shown in panels (b). Indices 1, 2 and 3 represent $\varphi=0, \pi$ and $\pi/2$, respectively.}
		\label{fig9}
	\end{center}
\end{figure}

The first scenario is represented by the pair $(m_1,m_2)=(1,2)$. In this case, there is no secondary shearless curves for $\varphi=0$ but the inclusion of a phase $\varphi\ne 0$ leads to the emergence of shearless curves. In Fig. \ref{fig9}, the internal winding number profile and the respective phase spaces are shown for $K_1=0.1$, $K_2=0.15$ and the three different values of $\varphi$.

As stated before, there is no shearless curve for $(m_1,m_2)=(1,2)$ and $\varphi=0$, as shown in Fig. \ref{fig9}(a1), where the winding number profile does not exhibit any extremum point. We observe the predominance of $m_2=2$ mode in the system with the existence of two islands, one around $x=0$ and the other around $x=0.5$. Different {configurations} emerge when the phase shift is different. For $\varphi=\pi$, we show the winding number profile and the phase space in Fig. \ref{fig9}(a2) and \ref{fig9}(b2), respectively. In this case, we observe a maximum point in the $\omega_{\mathrm{in}}$ profile, representing the only shearless curve {found} in the phase space. This bifurcation corresponds to the first pattern observed in Ref. \cite{leal2025}, where the single shearless curve emergence is related to the pitchfork bifurcation that occurs for the elliptic point. As verified by the phase space, we have the predominance of mode $m_2=2$.

For $\varphi=\pi/2$, we observe the winding number profile and the respective phase space in Fig. \ref{fig9}(a3) and \ref{fig9}(b3). For this case, we observe the emergence of a pair of shearless curves, represented by the pair of maximum-minimum local points in the $\omega_{\mathrm{in}}$ profile. This case represents the second pattern observed in Ref. \cite{leal2025}. Differently, for Fig. \ref{fig9}(b3) we have the predominance of mode $m_1=1$.

In summary, while no shearless curve is observed for $\varphi=0$, we observe the emergence of a single curve when $\varphi=\pi$ and the emergence of a pair of curves when $\varphi=\pi/2$. This difference occurs because of the different kind of bifurcation that occurs in the interior of the islands. For $\varphi=\pi$, we observe a pitchfork bifurcation, while for $\varphi=\pi/2$, as stated in Table \ref{tab1}, there occurs a saddle-node bifurcation.

\begin{figure}[!h]
	\begin{center}
		\includegraphics[width=0.5\textwidth]{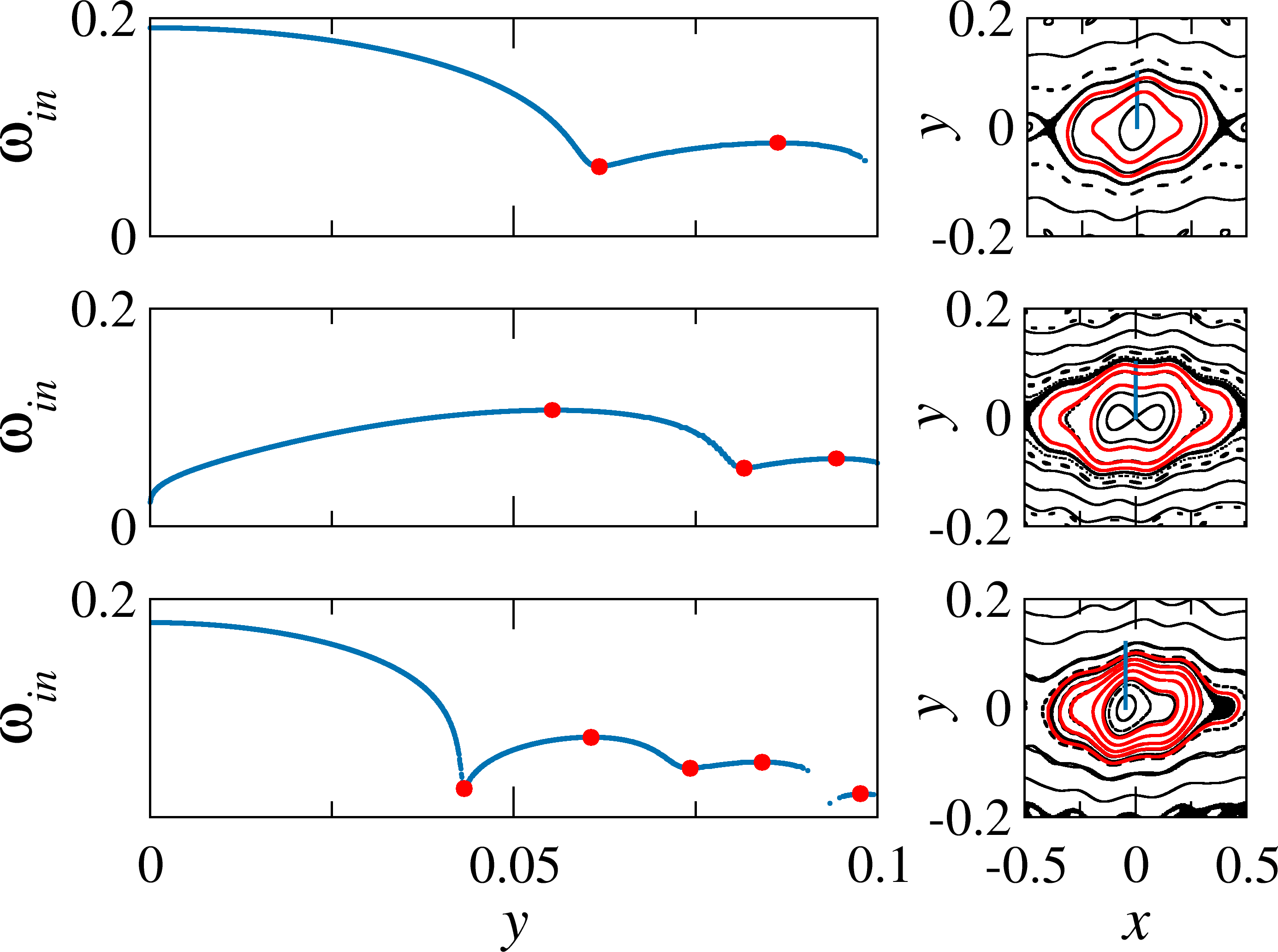}		
		\caption{Emergence of pairs of shearless curves for the two-harmonic standard map with and without phase shift. We chose $m_1=1, m_2=4$, $K_1=0.1$ and $K_2=0.25$. The internal winding number profile [panels (a)] shows pairs of maxima and minima, where each one represents a pair of shearless curves, displayed by the red curves in the phase spaces [panels (b)]. Just as in Fig. \ref{fig9}, each index indicates a different phase shift $\varphi$.} 
		\label{fig10}
	\end{center}
\end{figure}

Now, we analyze the effect of a non-zero phase shift which leads to a pair of shearless curves for $\varphi=0$. This scenario is illustrated by the pair $(m_1,m_2)=(1,4)$ and the results on the winding number profiles and the respective phase spaces are shown in Fig. \ref{fig10}.

As observed in Fig. \ref{fig10}(a1), for $\varphi=0$ there is a pair of shearless curves, indicated by the red points in the $\omega_{\mathrm{in}}$ profile and by the red curves in Fig. \ref{fig10}(b1). With a non-zero phase shift $\varphi=\pi$, we have the results shown in Fig. \ref{fig10}(a2) and \ref{fig10}(b2). In this case, we have a maximum in the $\omega_{\mathrm{in}}$ profile, around $y \approx0.053$, followed by a maximum-minimum pair for greater values of $y$. The first maximum occurs for all winding number profiles with $\varphi=\pi$. This occurs because the fixed point $(0,0)$ undergoes  a pitchfork bifurcation for all values of $m_1$ and $m_2$ and, as seen in Ref. \cite{leal2025} and in Fig. \ref{fig9}(a2), a single maximum in related to the occurrence of a pitchfork bifurcation. Lastly, for $\varphi=\pi/2$, we observe multiple pairs of maximum-minimum points in the $\omega_{\mathrm{in}}$ profile in Fig. \ref{fig10}(a3). As a consequence, we observe multiple shearless curves in the phase space of Fig \ref{fig10}(b3). 

Just as in the previous case, while we have the predominance of mode $m_2=2$ for $\varphi=0,\pi$, Fig. \ref{fig10}(b1) and \ref{fig10}(b2), the predominant mode for $\varphi=\pi/2$ is mode $m_1=1$, as shown in \ref{fig10}(b3). The second observed shearless bifurcation is related to the emergence of pairs of shearless curves for all studied phase-shifts $\varphi$.

\begin{figure}[!h]
	\begin{center}
		\includegraphics[width=0.5\textwidth]{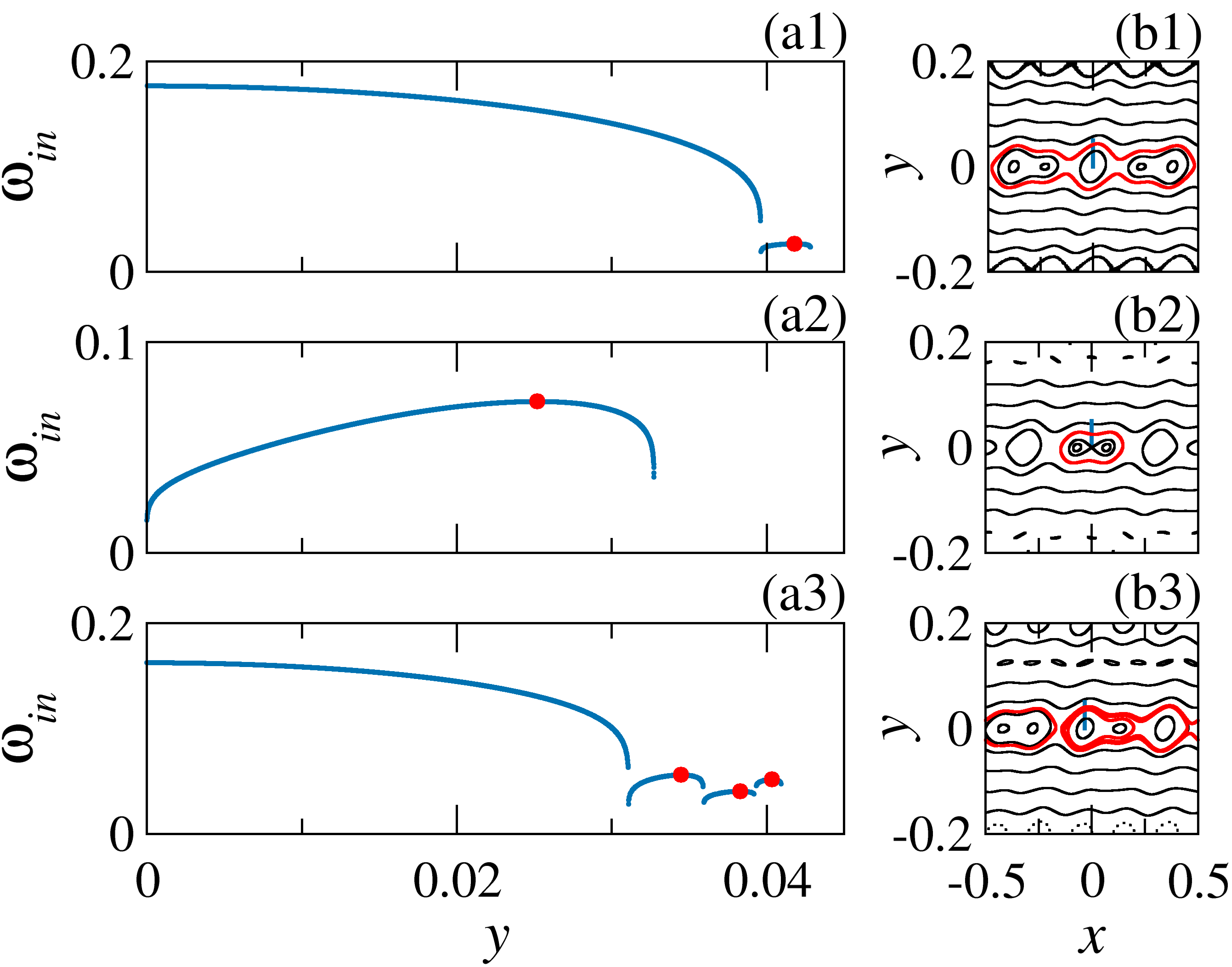}		
		\caption{Emergence of maximums points in the internal winding number profile. The parameters are $m_1=3, m_2=5$, $K_1=0.1$ and $K_2=0.2$. Each index represents a phase shift: (1)  $\varphi=0$, (2) $\varphi=\pi$ and (3) $\varphi=\pi/2$.}
		\label{fig11}
	\end{center}
\end{figure}

Lastly, we present the third observed scenario: the emergence of a maximum or multiple maximums in the internal winding number profile for all analyzed phase shift values. This scenario is represented by the pair $(m_1, m_2)=(3,5)$ and the respective results are shown in Figure \ref{fig11}.

The last observed bifurcation  includes the emergence of just maximum points in the $\omega_{\mathrm{in}}$ profile, as shown in panels (a) of Figure \ref{fig11}. For $\varphi=0$, we observe a single local maximum in Fig. \ref{fig11}(a1) corresponding to the only secondary shearless curve in Fig. \ref{fig11}(b1). A similar situation is observed in the second line, for $\varphi=\pi$, where only one maximum is present. A slightly different case is observed in the last line, for $\varphi=\pi/2$, where we observe multiple maximums in Fig. \ref{fig11}(a3). Each maximum represents a shearless curve, indicated by the red curves in Fig. \ref{fig11}(b3). For all phase spaces shown in Fig. \ref{fig11} (b), we have the predominance of mode $m_2=5$ with five distinct islands in the phase spaces.

\section{Conclusions}
\label{secVI}

Nonlinear coupling is a phenomenon widely studied in physical systems. The presence of a phase shift in such systems can lead to distinct bifurcation scenarios. In this research, the inclusion of a nonzero phase shift in the two-harmonic standard map was analyzed. We considered two synchronous modes in the map and three values for the phase shift: null phase shift, $\varphi=\pi$ and $\varphi=\pi/2$. From the phase space analysis, we observe distinct bifurcations. For $\varphi=\pi$, we observe the interchange of stability of the fixed points while, for $\varphi=\pi/2$, there are no fixed points in the usual position $x=0$ and $x=0.5$.

The role of the phase shift is crucial to the isochronous bifurcations of the system. We observe that {a nonzero phase shift makes it possible} to have different intermediate modes in the bifurcations, including the emergence of new intermediate modes in scenarios where there were not intermediate modes for null phase shift. However, for such values of $m_1$ and $m_2$, {e.g.} $m_1=4$ and $m_2=5$, an invariance in the modes and in the bifurcation lines is also a possible scenario.

The routes from mode $m_1$ to mode $m_2$ occur through pitchfork and saddle node bifurcations. A non-zero phase shift can alter the type of bifurcation that occurs to take the system from one mode to the other. We observe that, for $\varphi=\pi$, we have the predominance of pitchfork bifurcations while the saddle-node is more common for $\varphi=\pi/2$.

Lastly, we identified different scenarios for the emergence of secondary shearless curves inside the islands in the phase space. The first scenario is the appearance of a shearless curve, for $\varphi=\pi$, and a pair of curves, for $\varphi=\pi/2$, in a case where there are no shearless curves for the null phase shift. For $\varphi=\pi$, there is also the emergence of a single shearless curve, represented by a maximum in the internal winding number profile. It was observed that, when there is the emergence of a pair of shearless curves for the null phase shift, there is also the emergence of a pair for non-null phase shift. A similar scenario is observed where there is a unique shearless curve for the null phase shift: there is also the emergence of shearless curves in the phase space for non-null phase shift, but the emergence occurs as one at a time.

{We have three parameters that continuously vary: $K_1$, $K_2$ and $\varphi$. A complete description of the system is contained in a three-dimensional parameter space, with also bifurcation diagrams as a function of $\varphi$. In this paper, with the current sections with three $\varphi$'s, we give a first idea of the rich phase diagram.}

\section{Acknowledgments} 

This research received the support of the Coordination for the Improvement of Higher Education Personnel (CAPES) under Grant No. 88881.895032/2023-01, the National Council for Scientific and Technological Development (CNPq - Grant No. 301019/2019-3, 403120/2021-7, 443575/2024-0, 302665/2017-0) and Fundação de Amparo à Pesquisa do Estado de São Paulo (FAPESP) under Grant No. 2024/03570-7 and 2024/05700-5 . We would also like to thank the 105 Group Science \cite{105GS} for fruitful discussions.

\section*{Data availability}
The source code and data are openly available online in the Oscillations Control Group Data Repository \cite{OCG}.

\renewcommand{\appendixpagename}{Appendix}
\appendix
\section*{Appendix: Fixed point analysis}

\label{appendix}

For the map described in equation (\ref{eq:ESNM}), we have a fixed point of period 1 when {$y=0$ and}
\begin{eqnarray}
    \dfrac{K_1}{2\pi m_1} \sin(2\pi m_1 x_n)+\dfrac{K_2}{2 \pi m_2} \sin(2\pi m_2 x_n + \varphi)=P,
\end{eqnarray}
{with $P\in\mathbb{Z}$ for all orbit on the unit torus. When one considers $0 \le K_{1,2} \le 4$,  it occurs that $(K_1/m_1)+(K_2/m_2)< 2\pi$ for any choice of $m_2>m_1>0$. Thus, we set $P=0$. The so called ``accelerator modes" with $|y_{n+1}-y_n|\ge 1$ are possible only for $(K_1/m_1)+(K_2/m_2)\ge 2\pi$.}

Analyzing the three values of the phase we studied, we have the following. For $\varphi=0$, a fixed point of period 1 satisfies the equation
\begin{eqnarray}
    \dfrac{K_1}{m_1}\sin(2 \pi m_1 x^*)+\dfrac{K_2}{m_2}\sin(2 \pi m_2 x^*)=0
\end{eqnarray}
Thus, $(x^*,y^*)=(0,0)$ and $(x^*,y^*)=(0.5,0)$ are a fixed point for any pair $(m_1,m_2)$, since $m_{1,2}$ are integers. The same fixed points are found for $\varphi=\pi$ since the equation
\begin{eqnarray}
    \dfrac{K_1}{m_1}\sin(2 \pi m_1 x^*)-\dfrac{K_2}{m_2}\sin(2 \pi m_2 x^*)=0
\end{eqnarray}
is valid for $(x^*,y^*)=(0,0)$ and $(x^*,y^*)=(0.5,0)$ for any pair $(m_1,m_2)$. 

The case $\varphi=\dfrac{\pi}{2}$ is distinct, as we cannot generically find the fixed points, for arbitrary values of $m_{1,2}$. A fixed point for $\varphi=\pi/2$ satisfies the equation
\begin{eqnarray}
    \dfrac{K_1}{m_1}\sin(2 \pi m_1 x^*)+\dfrac{K_2}{m_2}\cos(2 \pi m_2 x^*)=0.
\end{eqnarray}
Therefore, we need to analyze the equation for each pair ($m_1,m_2$). We determine the fixed points for the pairs $(m_1,m_2)=(2,3)$, $(m_1,m_2)=(2,5)$ and $(m_1,m_2)=(4,5)$. For these three pairs, the fixed points are $(x^*,y^*)=(1/4,0)$ and $(x^*,y^*)=(3/4,0)$.\\


The eigenvalues for a fixed point of period 1 are
\begin{eqnarray}
\begin{aligned}
    \lambda=&\dfrac{-(K_1 \cos(2 \pi m_1 x^*)+K_2 \cos(2 \pi m_2 x^* + \varphi) -2)}{2} \hspace{1em}\pm \\
    &\hspace{10em}\dfrac{\sqrt{[(K_1 \cos(2 \pi m_1 x^*)+K_2 \cos(2 \pi m_2 x^* + \varphi) -2)]^2 -4}}{2}.
    \label{eign}
\end{aligned}
\end{eqnarray}
The value of $\lambda$ and, consequently, the type of fixed point (elliptic or hyperbolic) depend on the parameters of the system $K_1$, $K_2$, $m_1$ and $m_2$. The bifurcation occurs when the term under the square root is equal to zero, \textit{i.e.,} when the fixed point changes its stability: the real eigenvalues become complex or vice versa.

Using the fixed point values in (\ref{eign}), we obtain the bifurcation lines shown in Table \ref{tab2}.

\begin{table}[h!]
\begin{tabular}{|c|c|c|c|}
\hline
& & & \\
Line & Fixed point & $\varphi$& $m_{1,2}$ values \\
& & & \\
\hline
& & & \\
 \multirow{7}{*}{$K_2=4-K_1$}& $(0,0)$ & \multirow{2}{*}{$\varphi=0$} & Any value of $m_1$ and $m_2$ \\
 & $(1/2,0)$&  & Even $m_1$ and $m_2$\\ 
 & & & \\
 & $(1/2,0)$& $\varphi=\pi$ & Even $m_1$ and odd $m_2$\\
 & & & \\
 & $(3/4,0)$&$\varphi=\pi/2$ & $m_1=4$, $m_2=5$\\
 & & & \\
\hline
& & & \\
 \multirow{12}{*}{$K_1=K_2$}& $(1/2,0)$& $\varphi=0$ & $m_1$ and $m_2$ with different parities\\
 & & & \\
   & $(0,0)$& \multirow{2}{*}{$\varphi=\pi$} & Any value of $m_1$ and $m_2$ \\
 & $(1/2,0)$&  & $m_1$ and $m_2$ with same parities\\
 & & & \\
 & $(1/4,0)$&\multirow{5}{*}{$\varphi=\pi/2$} & \multirow{2}{*}{$m_1=2$, $m_2=3$}\\
  & $(3/4,0)$& & \\
  & & & \\
 & $(1/4,0)$& & \multirow{2}{*}{$m_1=2$, $m_2=5$}\\
 & $(3/4,0)$& & \\
 & & & \\
& $(1/4,0)$& & $m_1=4$, $m_2=5$\\
& & & \\
  \hline
\end{tabular}
\caption{General bifurcation lines for the two-harmonic standard map with phase $\varphi$. We identified two general bifurcation lines, indicated in the first column. In the second column, we present the fixed points which undergo the respective bifurcation. The phase is written in the third column while the conditions on $m_1$ and $m_2$ for the bifurcation to happen are indicated in the fourth column }
\label{tab2}
\end{table}

\end{document}